\documentclass[aps,12pt]{article}
\usepackage[tbtags]{amsmath}
\usepackage{amssymb}
\usepackage{amsfonts}
\usepackage{mathbbold}
\usepackage{hyperref}
\usepackage{txfonts}% \usepackage{mathptmx}
\usepackage{esint}
\usepackage{epic}
\usepackage{eepic}
\usepackage{times}
\usepackage[matrix,arrow]{xy}
\usepackage{epsfig}
\usepackage{slashed}
\usepackage[numbers,sort&compress]{natbib}
\usepackage[amsmath,thmmarks]{ntheorem}
\usepackage{mathrsfs}
\DeclareMathSizes{12}{13}{9.1}{6.5}
\def \ignore#1 { {} }

\def \Fig#1#2#3 {
\begin{figure}
\begin{center}
\scalebox{.6}{\includegraphics{#1.eps}} \label{#1}
\end{center}
\caption{#3}
\end{figure}
}

\newcommand{\figscaled}[3]{
\begin{figure}[t]
\begin{center}
\includegraphics[#3]{#1}
\end{center}
\caption{#2} \label{#1}
\end{figure}}

\topmargin by -\headsep \textheight 9.2in \oddsidemargin -25pt
\evensidemargin \oddsidemargin \marginparwidth 1.0in \textwidth
6.8in
\def \be {\begin{equation}}
\def \ee {\end{equation}}
\def \bea {\begin{eqnarray}}
\def \eea {\end{eqnarray}}
\def \beaa {\begin{eqnarray*}}
\def \eeaa {\end{eqnarray*}}

\def\bra#1{\langle{#1}|}
\def\ket#1{|{#1}\rangle}
\def \d {\rm d}

\newcommand{\mathsym}[1]{{}}

\theoremstyle{plain}

\newtheorem{propn}{Proposition}

\theoremstyle{remark}

\begin{document}

\title{AGT conjecture and AFLT states:  a
complete construction}

\date{}\maketitle
\begin{center}
\renewcommand{\thefootnote}{\fnsymbol{footnote}}

\centerline{Bao Shou \footnote{bsoul@itp.ac.cn},
Jian-Feng Wu \footnote{wujf@itp.ac.cn} ,
and Ming Yu\footnote{yum@itp.ac.cn}}

\vskip .5cm
\emph{
Institute of Theoretical Physics,\\\vspace{0.1cm} Chinese Academy of Sciences,
 Beijing, 100190, China}
\end{center}
 {\begin{abstract}
A complete construction of the AFLT states is proposed. With this construction and for all the cases we have checked, the AGT conjecture on the equivalence of Nekrasov Instanton Counting (NIC) to the $Vir\oplus u(1)$ conformal block has been verified to be true.
\end{abstract}}
\vspace{0.5cm}
\begin{flushleft}
\hspace{1cm}{\small PACS: {11.25.Hf, 12.60.Jv, 12.40.Nn, 02.30.Ik}}
\end{flushleft}

\begin{flushleft}
\hspace{1cm}{\small{\bf Keywords:}{\ Liouville Theory, Conformal Blocks, AGT Conjecture, Supersymmetric Gauge Theory,\\\hspace{2.7cm} S-duality}}
\end{flushleft}

\newpage

\section{Introduction}
Conformal blocks, which are defined on the
(punctured) Riemann surfaces, holomorphic in each $z_i$ coordinate except when they meet each other, play an essential role in building correlation functions in two dimensional (Euclidean) conformal field theories\cite{BPZ}. They can be best understood as sewing together chiral vertex operators\cite{Sonoda:1988mf,Sonoda:1988fq,Moore:1988qv}, which by definition, are not  local objects, but the
correlation functions are. The later combine both holomorphic and anti-holomorphic conformal blocks in a consistent way to make modular covariant objects. On the sphere, the $n$-point conformal
block is represented graphically as in fig.\ref{confblock}, where $h_i$ is the conformal dimension of the primary field inserted at coordinate $z_i$, and
$\tilde{h_i}$ labels the contribution arising from the conformal family descending from a primary field
with the conformal dimension $\tilde{h_i}$. The global conformal invariance is $SL(2) \times SL(2)$, which may be used to fix three  coordinates
$z_1=0$, $z_{n-1}=1$ and $z_n= \infty$. So the independent variables are $z_i$, $i=2,...,n-2$, with
the degrees of freedom $n-3$ for the $n$-point conformal blocks on the sphere.

The calculation of conformal blocks is based on the conformal Ward-identities,
$$[L_n, V_h(z)]=(z^{n+1}\partial_z +(n+1)h z^n )V_h(z).$$ and  carried out perturbatively level by level \cite{BPZ,Zamolodchikov82,Zamolodchikov84}. In some special cases, the decoupling of the Virasoro null vectors can be implemented as differential equations for the conformal blocks. For the general case,  recursion relations have been proposed by Zamolodchikov\cite{Zamolodchikov82,Zamolodchikov84}
on the meromorphic structures of the conformal blocks either in complex $c$-plane or $h$-plane. However, in general,  the global perspective of the sewing procedure for the conformal blocks was still not fully understood until recently when the AGT duality \cite{AGT} had been proposed.

AGT conjecture relates 2d Liouville conformal field theories to 4d $N=2$ supersymmetric gauge theories of the $A_1$ type. The main idea is coupling to the Liouville field a $u(1)$ field\footnote{In fact, the zero mode of the $u(1)$ field is a gauge symmetry and can be fixed to any desired value.}, then this system is dual to a $U(2)=SU(2) \times U(1)$ superconformal 4d theory. In this case,
the partition function by Nekrasov instanton counting(NIC)\cite{Nekrasov:2002qd,Nekrasov:2003rj} of the 4d $U(2)$ theory is to be identified with the conformal blocks of the $u(1)$ coupled Liouville type. The Liouville CFT is characterized by a 2d one boson theory with center charge $c\ge 25$. Finally, one can decouple the $U(1)$ factor and obtain the
instanton partition function of the $SU(2)$ theory which duals to Liouville
conformal blocks.
Liouville interaction breaks down the charge conservation explicitly and leads to the introduction of the screening charges. Because of the existence of the screening charges, the conformal blocks of the Liouville type is much more complicated than its counterpart
of the $u(1)$ free boson theory. However, the AGT conjecture, if proven true, means that there exists an orthogonal basis upon which the $Liouville\times u(1)$ conformal
blocks are built. From the above reasoning, there exists a tree-like structure which describes the duality in coupling space of the $N=2$  4d superconformal linear quiver gauge theory. The primary objects for this tree-like structure is the ``bifundmental'' matter coupling, which, if translated correctly, should be represented by the inner products of the bra and ket descendant fields in 2d conformal families sandwiched by a ``primary'' vertex operator at position, say, $z$. Such kind of pants-like diagram can be sewed together to form a linear quiver diagram, which, on the 2d CFT side,  is
just the $n$-point functions on the sphere for our consideration. Of course, in the present context, we mean the $Vir\oplus u(1)$ 2d CFT.

At first sight, it seems that such duality does not bring in any conveniences.
However, the Nekrasov instanton counting on the 4d field theory
shows a rather compact form for the summands which are completely factorized in ``momentum'' $P$. And the summation is well organized into the combinatorial enumeration of the Young tableuax. This simple
structure implies Liouville theory, in particular, the evaluation of the Liouville conformal blocks,  could be resolved by embedding it
into a bigger system. So one may expect a new construction for the
Liouville conformal blocks from the corresponding NIC.
\figscaled{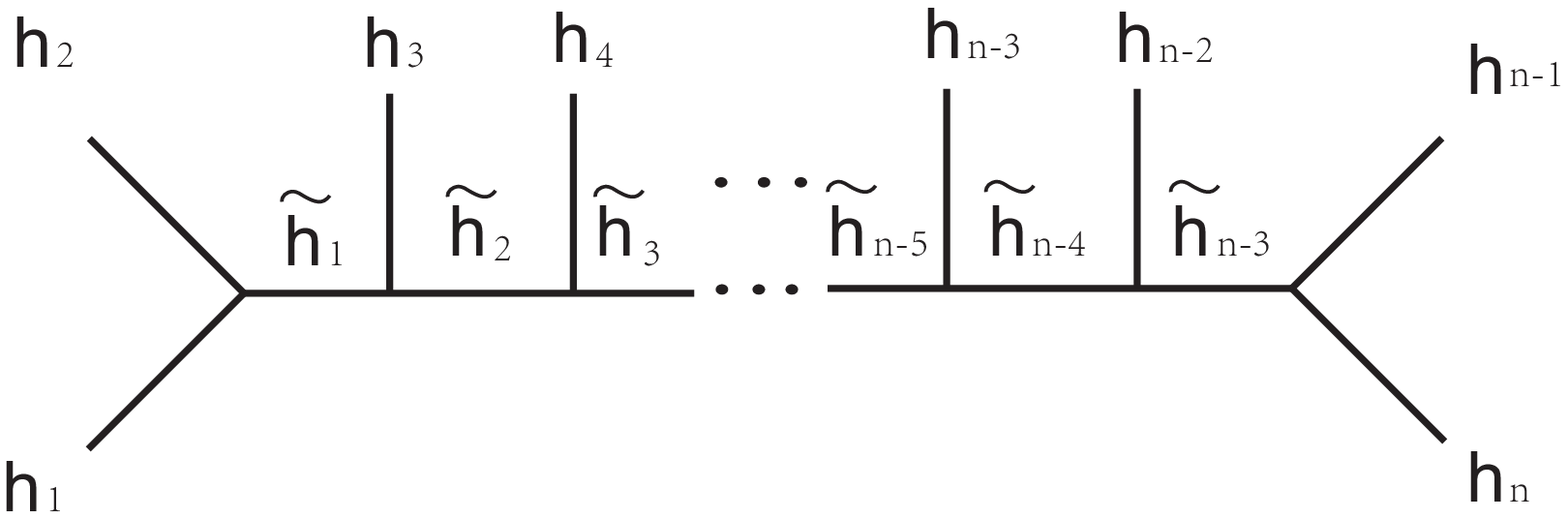}{$n$-point conformal block on $S^2$}{height=1.5in}

As pointed out by Nakajima\cite{Nakajima:2003pg,Nakajima:2005fg,Nakajima:2003uh}, the instanton counting for $N=2$ gauge theory is equivalent to the Hilbert scheme of points on the corresponding Seiberg-Witten curve (blow-up Riemann surface)\cite{Seiberg:1994rs,Seiberg:1994aj,Witten:1997sc}. This can be translated into a topological string description from physicists' point of view.  By invoking the D4-D0 brane setup\cite{Douglas:1995bn,Seiberg:1999vs} for ADHM construction\cite{Atiyah:1978ri} of the instanton moduli space and the resolving process for ALE singularities\cite{Katz:1996fh}, these indicate that the instanton counting is a counting for D0 branes in a toric Calabi-Yau 3-fold.  Actually, there are two kinds of D0 branes in the Calabi-Yau 3-fold, one is the regular D0 brane, which is in regular representation of $\Gamma$, the center of the corresponding ADE group. It carries no flux and can move freely on the Riemann surface. The other is the fractional D0 brane, which is a D2 brane wrapping on a zero-sized two sphere. It is always attached to the ALE singularity since it has a nontrivial monodromy while moving around the singularity. It is these fractional D0 branes that resolve the ALE singularity, and leave fluxes on the blow-up Riemann surface. This property ensures that one can identify these fractional instantons as ``anyons'' on the Riemann surface. On the other hand, the regular ones are ``electric charged'' particles on the Riemann surface. So the total counting is equvalent to solving the problem of ``electron gas'' system with insertions of anyons at the blow-up singularity on the Riemann surface. This point of view is partialy included in Dijkgraaf and Vafa's article\cite{Dijkgraaf:2009pc}. For each pants of the pants decomposition for the (punctured) Riemann surface, one can guess that the instanton partition function can be rewritten as summation over all the intermediate states passing through the sewn holes\cite{Moore:1988qv}. For the interests of the present paper, we concern ourselves only with the special pants diagram that one of the tubes is replaced by the blow-up singularity. Then the summand in the instanton partition function represents itself as an 
inner product of the bra and ket states, sandwiched by the anyonic vertex operator. These bra and ket states should come from the interacting ``electronic''{\footnote{For each simple root of an ADE group, one should introduce a kind of ``electronic'' field.}} particles. A candidate description of the ``electronic gas'' system is the integrable system of multiple Calogero-Sutherland model, each living on a cycle. The whole (punctured)  Riemann surface, can be obtained by sewing together these pants on nonintersecting cycles.

There are many efforts on relating the conformal blocks to the NIC\cite{Nekrasov:2009rc,Alday:2010vg,Cheng:2010yw,Itoyama:2010ki,Belavin:2011js,Belavin:2011pp,Dorey:2011pa,Awata:2010bz,Mironov:2010su,Mironov:2010zs,Wu:2010rh} from various points of views, and these works confirm the validity of the AGT duality. However, the explicit construction for the Liouville conformal blocks has remained  largely unclear until the recent work \cite{AFLT}
by Alba, Fateev, Litvinov and Tarnopolsky. In \cite{AFLT}, they
have put forward the AGT duality in a more explicit form \bea
\label{AFLT}\dfrac{_{\vec{Y}'}\bra{P'}V_{\alpha}\ket{P}_{\vec{Y}}}{\bra{P'}V_{\alpha}\ket{P}}
&=& Z_{bif}(\alpha|P',\vec{Y}'; P,\vec{Y}) \, \, ,
 \eea
 here specifically for a free field realization, \[V_{\alpha}(z) =
e^{2i(Q-\alpha)\tilde{\varphi}_-(z)}e^{-2i\alpha\tilde{\varphi}_+(z)}S^n:e^{2i\alpha\varphi(z)}:\,,\] with $P + P' +\alpha +nb=0$\, , and $S=\oint e^{2ib\varphi(z)}\d z$ is the screening charge in the Virasoro sector.
The l.h.s. of eq(\ref{AFLT}) is the  pants-like (with one of the
tubes labeled by $\alpha$ shrinks to a line) conformal block. The r.h.s. of
(\ref{AFLT}) reproduces $Z_{bif}$ for the instanton counting, which is
given by \bea\label{zbif}Z_{bif}(\alpha|P',\vec{Y}'; P,\vec{Y}) =
\prod_{i,j=1}^2 \prod_{s\in Y_i}\left(Q-E_{Y_i,Y_j'}(P_i - P_j'|s) -
\alpha\right)\prod_{t\in Y_j' }\left(E_{Y_j',
Y_i}(P'_j-P_i|t)-\alpha\right)\,,\eea where $\vec{P} = (P,-P)$,
$\vec{P'} = (P',-P)$ and \bea E_{Y,Y'}(P|s) \equiv
P+b^{-1}(a_{Y}(s)+1)-bl_{Y'}(s) \,. \eea Here $a_Y(s)$ and
$l_{Y}(s)$ resp. are the arm length and the leg length resp. of the box $s$ in the
Young tableau $Y$, defined as
\[a_{Y}(s)|_{s=(i,j)} := \lambda_i - j,\ \ \ \  l_{Y}(s)|_{s=(i,j)} := \lambda_j^t -i
\,,\] $\lambda_i$ and $\lambda_j^t$ resp. are  the $i$-th
part of the partition $\lambda = (\lambda_1, \lambda_2, \cdots) ,
\,\, \lambda_i \geq\lambda_{i+1}$ and the $j$-th part of the transpose
partition $\lambda^t$ respectively .

 (\ref{AFLT}) means that the matrix elements of a special ``chiral
 vertex operator'' $V_\alpha$ in a suitably chosen basis,  can be translated
  into a 4d theory as an instanton contribution for a special bifundamental
  contribution of the NIC. By sewing together pants-like diagrams one gets any
  desired duality diagrams in the coupling space of the linear quiver gauge theory. 
  So, the checking of the AGT
  duality reduces  to the construction of the states $\ket{P}_{\vec{Y}}$,  which we shall call the AFLT states\cite{AFLT}, with $\vec{Y}\equiv (Y_1,Y_2)$ the Young tableaux.
Here the $Y$'s, the partitions of natural numbers, or equivalently represented
 by  Young tableaux, are labels for the orthogonal basis for the descendant
 fields (Verma modules) in a $Vir\oplus u(1)$ conformal family from the 2d CFT point of view.
By definition, the AFLT states form a complete set of states for the family members in a given $Vir\oplus u(1)$ conformal family and  the inner products between them, sandwiched by a vertex operator of the particular form, $V_\alpha(z)$, at position,say, $z=1$,
is factorized exactly as the NIC $Z_{bif}$ presented on the r.h.s. of (\ref{AFLT}).
 The explicit formula, (\ref{AFLT}), puts strong constraints on the possible forms of the AFLT states and make a systematic construction of them
unaccessible at first glance. In \cite{AFLT}, only the explicit form of the state $\ket{P}_{Y,\varnothing} $ has been found, $$\ket{P}_{Y,\varnothing} =J^+_{-Y}\ket{P}\Omega_Y(P),$$ with   $J^+_{-Y}$ the creator $(-ib)^{-1}a^+_{-n}$'s valued Jack symmetric function, and $\Omega_Y(P)$ the normalization constant.

In our opinion, the AGT conjecture, written in the form of (\ref{AFLT}), strongly suggests that the $Vir\oplus u(1)$ conformal family is a Hamiltonian system with $\ket{P}_{\vec{Y}}$ the Hamiltonian eigenstates. So the construction of the AFLT states becomes a quantum mechanical problem of solving the Schrodinger equation.
Put things in this way, we propose a possible form of the Hamiltonian $H$ and construct its eigenstates explicitly. We shall identify those eigenstates as the AFLT states desired. For in all the cases we have checked, (\ref{AFLT}) is verified to be true, using the AFLT states we have constructed.
  We shall present now as the main results of our present paper the explicit form of the Hamiltonian $H$ along with the complete construction of the AFLT states, $\ket{P}_{\vec{Y}}$. More elaborated exposition will come in the subsequent sections.
\bea\label{main1}
H&=&H_0+H_I \\\nonumber
\ket{P}_{\vec{Y}}&=&\frac{1}{1-\frac{1}{E_{\vec{Y}}(P)-H_0}H_I}J_{-\vec{Y}}|P\rangle \Omega_{\vec{Y}}(P).\eea
Here, $J^\pm_{\pm Y}$ are the Jack states constructed  in terms of the oscillators $a^\pm_n$'s or $a^\pm_{-n}$'s ($n>0$) solely, $H^\pm$  the corresponding Hamiltonian for the Jack symmetric functions,  $H_0\equiv H^+ +H^-$. Thus the eigenstate of $H_0$ is just  $J_{-\vec{Y}}\ket{P}\equiv J_{-Y_1}^+J_{-Y_2}^-\ket{P}$ with the eigenvalue $E_{\vec{Y}}(P)$.
$H^\pm$ in our formalism is defined to include zero modes $a^\pm_0$ also, $-i a^\pm_0\ket{P}=\pm P\ket{P}$.
It is important that $H_I$ is strictly triangular with respect to the basis vectors of the $H_0$ eigenstates. By ``strictly triangular'' we mean the (upper or lower) triangular matrix with zero diagonal entries. It is easy to see that if the interaction term $H_I$ is strictly triangular, then the eigenvalue spectrum of $H_0$ remains unperturbed and $\ket{P}_{\vec{Y}}$ in (\ref{main1}) well defined for non-degenerate $H_0$ spectrum descending from a mother state $J_{-\vec{Y}}\ket{P}$ for generic values of $P$'s.
Putting things all together, we have
\bea\label{main2}
&&H=H_0+H_I,\ \ \ H_0= H^+ +H^-,\ \ \ H_I=\sum_{n=1}^{\infty}2Qn a_{-n}^+a_n^-,\\\nonumber
&&H^{\pm}=\dfrac{-i}{3}\oint\left(z\partial_z\varphi^{\pm}\right)^3 \frac{dz}{2\pi i z}
+\sum_{n=1}^{\infty}Q n a_{-n}^{\pm}a_n^{\pm},\\\nonumber
&&E_{\vec{Y}}(P)=E_{Y_1}+E_{Y_2}+2P(|Y_1|-|Y_2|),\ \ \ E_Y=\sum_i (y^{2}_{i}b^{-1}+(2i-1)y_i b),\\ \nonumber
&&\Omega_{\vec{Y}}(P)=
(-)^{|Y_1|}b^{|Y_1|+|Y_2|}\prod_{Y_1}\left(2P+(a_{Y_1}+1)b^{-1}-l_{Y_2}b\right)
\prod_{Y_2}\left(2P-a_{Y_2}b^{-1} +(l_{Y_1}+1)b\right),\\\nonumber
&&\ket{P}_{\vec{Y}}=\frac{1}{1-\frac{1}{E_{\vec{Y}}(P)-H_0}H_I}J_{-\vec{Y}}|P\rangle \Omega_{\vec{Y}}(P),\\\nonumber
&&H_0J_{-\vec{Y}}\ket{P}=E_{\vec{Y}}(P)J_{-\vec{Y}}\ket{P},\ \ \ H\ket{P}_{\vec{Y}}=E_{\vec{Y}}(P)\ket{P}_{\vec{Y}},\ \ \ -i a^\pm_0\ket{P}=\pm P\ket{P}
\eea

Notice that \\
1) $\ket{P}_{Y,\varnothing} $ constructed in \cite{AFLT} are included in our construction as subcases.\\
2) The Hamiltonian $H$ constructed by us, albeit in a disguised form, turns out to coincide up to some trivial factor with $I_3$, one of the integrals of motion found in a different context in appendix C of \cite{AFLT}. $I_3$ in \cite{AFLT}, written in the form of $Vir\oplus u(1)$,
makes the Virasoro symmetry manifest, but is not suitable for solving a perturbation theory with
perturbation parameter $Q=b+b^{-1}$. The Hamiltonian $H$ written in terms of the interacting bi-Jack polynomial system as in (\ref{main2}), shows Virasoro symmetry only implicitly, but makes
the perturbation theory exactly solvable as we shall see soon after.

The procedure is outlined as follows. On the 2d CFT side, the $Vir\oplus u(1) $ theory can be represented as a theory of two independent scalars $\tilde{\varphi}(z)$ and $\varphi(z)$. $\tilde{\varphi}(z)$ part is essentially a free theory of timelike oscillators, while the scalar field $-i\varphi(z)$
 is spacelike but engaged in a Liouville type interaction. The two scalars can be linearly combined to form the ``light-cone'' scalars.
$\varphi^+(z)$ and $\varphi^-(z)$. The labeling $\vec{Y}$ of the basis vectors strongly suggests that there exist a bi-Jack polynomial structure, plus possibly some interactions between these two sectors. That is, the ``free'' $H^\pm$ spectrums should be described by $J_{Y_1}^{+}$ and $J_{Y_2}^-$ respectively, here $J_Y$ denotes Jack states related to Young tableau $Y$.  First we construct the ``unperturbed'' energy operator $H_0$
which just sums up the ``energies'' in $J^{\pm}_Y$ sectors, $H_0= H^+ +H^-$. The
next thing is to specify the interaction between these two sectors. Strictly speaking,
$H_0$ does not describe a free theory, since it also contains the interaction terms proportional to $Q$. But the new interaction term $H_I$ further mixes the $J^\pm_Y$'s
and the coupling is also a first order in $Q$. It is good to see that $H_I$ is strictly triangular with respect to the basis vectors of $H_0$ eigenstates.
Our method can be easily generalized to
wider classes of integrable models, in which the interacting Hamiltonian splits into two parts, $H^{//}$ and $H^{\perp}$, representing respectively the shift of energies and the rotations (mixings) of states. The later keeps the eigenvalue spectrum untouched\cite{Wu:2011dz}.

Besides being triangular, the form of the interaction term
is  however much restricted, also by the Virasoro symmetry. Since the total Hamiltonian is of the form $Vir\oplus u(1)$, an ``interaction energy operator'' $H_I$ is needed to make the ``full Energy operator'' $H=H_0+H_I$ the combination of $a_n$'s and $L_n$'s only. Once the Hamiltonian structure is determined, then the construction of the Hamiltonian eigenstate $\ket{P}_{\vec{Y}}$ is just a quantum mechanical problem. $H_0$ and $H$ share the same eigenvalue spectrum, but only the eigenstates of $H$, represented by $\ket{P}_{\vec{Y}}$'s, form a complete set of basis vectors for the $Vir\oplus u(1)$ conformal family.

We have checked by examples the corresponding AGT duality formula, (\ref{AFLT}) up to level 4, and 
have found that indeed Nekrasov instanton counting can be reproduced with this construction, (\ref{main2}). In fact, we have also checked more general cases and all get positive answers. But those more general  results will appear elsewhere due to lacking of space to include them in this paper.

The insertions of the screening charges play an important role in checking the AGT duality. However, in the present work we concern ourselves only with the cases in which the screening charges can be detached away from the vertex operator $V_\alpha$ and moved on to act on the AFLT states (similar to the Felder's BRST operators)\cite{Felder:1988zp,Bernard:1989iy}. The more general cases in which screening charges can not be moved away from $V_\alpha$ will be under our future studies.

It is well known that it is possible to map the Liouville theory to the analytic continuation of the Calogero-Sutherland(CS) model,
which was originally and in most cases considered to be a theory with the parameter  $\beta >0$,  while in the Liouville case $\beta <0$ is required. Some explanation is given in \cite{Wu:2011cy}.
The physical space of the
CS model are created by Jack polynomials, which are symmetric
functions studied in great detail in mathematics and physics
literatures\cite{Stanley,Macdonald}. The integrability of the CS model may be derived in different ways, e.g., from the knowledge of the hidden $W_{1+\infty}$ symmetry
of the model. A recursion relation related to the Virasoro singular vectors and an integral representation based on it has appeared recently in \cite{Wu:2011cy}, in which more references can be found on the subjects of the CS model and the Jack symmetric functions.  It should be stressed again that for $\beta >0$, there is no null vectors in the CS model. So the ``null'' vectors are not the true null vectors of the CS model, since the Virasoro algebra based on which the null vectors are constructed is not the true conformal algebra of the CS model  in that case.
But for $\beta <0$, yes, there are null vectors in the CS model. It is possible to describe the $Liouville \times u(1)$ theory in terms of the Jack polynomials considered as analytic continuation from $\beta>0$ to $\beta<0$.

There is another hint that the $Liouville \times u(1)$ theory has something to do with $\beta <0$ CS model. It can be found from the Nekrasov partition function, in which each term in the summation
can be written in the form of the Carlsson-Okounkov formula\cite{Carlsson-Okounkov},
for the special cases when no screening charges are inserted.
Carlsson-Okounkov formula is a formula for the inner products between the bra Jack states and the ket Jack states sandwiched with a modified vertex operator. This
extraordinary formula is of great help in checking the AGT duality with our construction for the orthogonal basis vectors $\ket{P}_{\vec{Y}}$'s defined in (\ref{AFLT}).

We notice that the construction we found shares many similarities with the construction of the Jack functions themselves.
Namely,
we take the state $J_{-Y_1}^+J_{-Y_2}^-\ket{P}$ as the mother state and
its descendants are constructed in such a manner that two partitions are ``squeezed''
into other pairs. The squeezing does not change the
total level of the two partitions, but does make the inner products of the descendants a triangular form.

Although the 4d to 2d duality has  just begun to be understood, it has been known for sometime that 2d conformal blocks can be equivalently described as insertions of Wilson
lines in 3d pure Chern-Simons topological gauge theory. In fact, we can
interpret the $n$-point conformal block represented by fig.1 as a Wilson line insertion inside a three-ball. The path integral in Chern-Simons-Witten gauge
theory thus creates a state living on the boundary of the three ball, which is punctured $S^2$. So it should not be a too big surprise that 2d conformal field theory has something to do with higher dimensional quantum field theories. Taking into account that Jack symmetric polynomials can be taken as some special limit of the two parameter Macdonald symmetric polynomials, one natural guess is that our construction can be generalized to the case of Macdonald symmetric polynomials. In that case there should be a 5d to 3d duality.

This paper is organized in the following way. Our general formalism on the construction of the AFLT states is presented in the introduction. In section 2, we explore the general structure of the $Vir\oplus u(1)$ conformal family. We found in some cases it is more convenient to work with the bi-Jack function basis. Section 3 contains the major derivation of our construction. Section 4 is the conclusion. And in appendix A 
the explicit construction of the AFLT states up to level 3 is presented.

\section{Exploring the $Vir\oplus u(1)$ Structure}
We are dealing with a 4d N=2 $U(2)$ linear quiver gauge theory coupled to special  bi-fundmental matter in a superconformal way. According to the standard AGT duality dictionary, the corresponding 2d conformal block is of the  $Vir\oplus u(1)$ type, which reproduces the instanton part of the Nekrasov partition function for the $U(2)$ theory. There are  two sets of Young
diagrams which measure the partitions of the instantons. If one wants to extract the Virasoro basis of the
conformal blocks, one need to factor out the $u(1)$ factor.

In this section we shall mainly explore the Hilbert space for the $Vir\oplus u(1)$ theory and find the requirements that the energy operator $H$ should meet. Our procedure depends heavily on the Nekrasov instanton counting formula written more suitably for the construction of the conformal blocks, (\ref{AFLT}). First, the 2d $u(1)$ conformal block, realized in terms of the oscillators of the scaler field $\tilde{\varphi}$, is essentially of free theory with center charge $c=1$. The zero modes can be integrated out trivially and does not play any significant role here. The vertex operators for  $\tilde{\varphi}$, take a peculiar form
$$e^{2i(Q-\alpha)\tilde{\varphi}_{(-)}(z)}e^{-2i\alpha\tilde{\varphi}_{(+)}(z)}$$
Here, $\tilde{\varphi}_{(\pm)}$ means the positive (negative) mode part of the scalar field
$\tilde{\varphi}$ .
Although the above vertex operator is not the  standard one in 2d CFT, its contribution to the conformal block can be easily read off and factored out.
Second, the $Vir$ part is a Liouville conformal field theory of the $\varphi (z)$ scalar field and is more complicated because of the existence of the screening charges.

We have the following mode expansion for the scalar fields $\varphi (z)$ and $\tilde{\varphi}(z)$,

\bea\label{scalars}
\varphi (z)&=&q+c_0 \log (z)+\sum_{n\in Z,n\neq 0}\frac{c_{-n} }{n}z^n \, ,\\\nonumber
\tilde{\varphi} (z)&=&\tilde{q}+a_0 \log (z)+\sum_{n\in Z,n\neq 0}\frac{a_{-n} }{n} z^n \, ,\\ \nonumber
\left[c_{n},c_{m}\right] &=& \frac{n}{2}\delta_ {n+m,0}\, ,\ \ \ \,(c_{-n})^{\dagger} = c_n\,,\ \ \ \left[c_0, q\right]=\frac{1}{2}  \\\nonumber
\left[a_{n}, a_{m}\right] &=& \frac{n}{2}\delta_{n+m,0}\, ,\ \ \ \,(a_{-n})^{\dagger} = -a_n\,,\ \ \ \left[a_0, \tilde{q}\right]=\frac{1}{2}\,  . \\\nonumber
\eea
Virasoro generators in the $Vir$ part, $L_n$,  thus reads \bea\label{OringinalLn} L_n
&=&\sum_{k\in\mathbb{Z}}c_{k}c_{n-k}-inQ c_n=
\sum_{k\neq0,n}c_{k}c_{n-k}+i(2\hat{P}-nQ)c_n,\\\label{L0}L_0 &=&
\frac{Q^2}{4}-\hat{P}^2+2\sum_{k>0}c_{-k}c_{k}\,,\eea here,
$c_0 = i\hat{P}$,  \[-ic_0\ket{P}=
\hat{P}\ket{P}= P\ket{P}, \,\,\,\bra{P}(-ic_0)=\bra{P}\hat{P} =
-P\bra{P}.\] 
By this construction, $L_n$
defined in (\ref{OringinalLn}-\ref{L0}) is obviously unitary, \[L_{-n}
= L_n^{\dagger}.\] In 2d CFT, one frequently meets another (more conventional) definition of the Virasoro generators, \be L^0_{n} = \sum_{k\in
\mathbb{Z}}c_{k}c_{n-k}- i Q(n+1)c_n\,.\ee
If (\ref{OringinalLn}) and ({\ref{L0}}) are combined in this way, \be\label{TZ1} T(z) = \partial\varphi\partial\varphi
+i Q \partial^2\varphi + i Q z^{-1}\partial\varphi - \frac{Q^2}{4}\frac{1}{z^2} = \sum_n L_n
z^{-n-2}\,, \ee then $T(z)$ differs from the more conventional one
$T^0(z)=\partial\varphi\partial\varphi + iQ \partial^2\varphi$ by a similarity
transformation \bea\label{TZ2} T(z) &=& e^{-iQ
q}(\partial\varphi\partial\varphi+iQ\partial^2\varphi)e^{iQq}=\sum_n \tilde{L}_n
z^{-n-2}\\\nonumber
\tilde{L}_{n} &=& \sum_{k\in
\mathbb{Z}}\tilde{c}_{k}\tilde{c}_{n-k}- i Q(n+1)\tilde{c}_n\,\,. \eea

Comparing (\ref{TZ1}) and (\ref{TZ2}), we have
\bea\tilde{c_n} =
\left\{\begin{array}{l}c_n,\,\,\, n\neq0\\c_0+\frac{i}{2}Q, \,\,\,
n=0\end{array} \right.\,.\eea

Viewing  the $Vir\oplus u(1)$  model as a 2d sigma model, since under conjugation $c_n$ and $a_n$ transform differently we recognize that $-i\varphi$ is spacelike and $\tilde\varphi$  timelike when they are considered as coordinates in target space. So the target space of the sigma model under consideration is curved in space and flat in time direction.
The two scalars can also be linearly combined to form the ``light-cone'' scalars $\varphi^\pm(z)$,
\bea\label{pm}
&& \varphi^\pm(z)= \tilde{\varphi}(z)\pm \varphi(z) \, , \\\nonumber
&&\varphi^\pm(z) \varphi^\pm(z')=\log (z-z')\, , \\\nonumber
&&\varphi^\pm(z) \varphi^\mp(z')=0\, , 
\\\nonumber
&&\varphi^{\pm \dagger}(z)=\varphi^\mp(z)\,.
\eea
The descendant states in the conformal family split into sub-spaces of different levels, which are measured by $I_2=L_0+ \sum_{n=1}^\infty a_{-n}a_n$. Within the sub-space of given level $N\equiv \sum_{n>0}(a_{-n}a_n+c_{-n}c_{n})$, states can be labeled either by linear conbinations of $a_{-X}L_{-Y}$'s or $J^+_{-X}J^-_{-Y}$'s, with $X$, $Y$ the Young tableau, $|X|+|Y|=N$, $J^{\pm}_{\pm Y}$ the annihilator $(-ib)^{-1}a_Y^{\pm}$'s (or creator $(-ib)^{-1}a_{-Y}^{\pm}$'s) valued Jack symmetric functions. In either case, one can infer from AGT duality that there exist a Hermitian operator $H$, which commutes with $I_2$  and diagonizes this subspace. Hence the eigenstates of $H$ form an orthogonal basis. We know $I_2$ acts on this subspace trivially like an identity operator. So in order to eliminate the degeneracy, the next candidate $H$ we are looking for should be at least cubic in the oscillators $a_{n}$'s and $c_{n}$'s. Once $H$ is introduced, the descendant states will organize themselves into an orthogonal basis labeled by two sets of Young tableaux $\{Y_1,Y_2\}$.
In our opinion, it is better to start with the $J^\pm_Y$
system, since there is already a Hamiltonian structure
$H^\pm$ acting separately on them. But $H_0=H^+ + H^-$ does not commute with the screening charges $S^\pm$ pertaining to $L_n$, $$S^\pm=\oint e^{2ib^\pm\varphi(z)}\d z \, ,$$
here $b^+\equiv b, \, b^-\equiv b^{-1}$ .
We then add a new term $H_I$ to $H_0$, $H=H_0+H_I$  and require that $\left[S^\pm, H\right]=0$. If $H_I$
are chosen correctly, the eigenstates of $H$ will coincide with the unique orthogonal basis $\ket{P}_{\vec{Y}}$, which we call AFLT states and are defined to satisfy (\ref{AFLT}), in which the matrix elements
$\frac{_{\vec{Y'}}\bra{P'}V_\alpha \ket{P}_{\vec{Y}}}{\bra{P'}V_\alpha \ket{P}}$ is factorized in a consistent way.

On the 4d theory side,  one can decouple a single massless bifundamental matter\footnote{The
massless condition implies $|Y'_1|+|Y'_2|=|Y_1|+|Y_2|$.} $(\vec{a} =
(P,P), m=0)$. We shall show that under this condition the contributions can be written as
the orthogonality condition for the $\ket{P}_{\vec{Y}}$'s, provided (\ref{AFLT}) is satisfied. 
\begin{propn}\footnote{This is actually Proposition 2.4 in \cite{AFLT}, but here we give more details.}
If AGT conjecture is true, then the AFLT states, $\ket{P}_{\vec{Y}}$'s defined in (\ref{AFLT}), form an orthogonal basis.
\bea\label{Bifund}
\!\!
 _{Y'_1,Y'_2}\langle  P\ket{P}_{Y_1,Y_2}&=& Z_{bifund}^{U(2) inst}(\vec{a},\vec{Y},\vec{a},\vec{Y'};0) \propto\delta_{\vec{Y},\vec{Y'}}.\eea
\end{propn}
\hspace{-0.6cm}\textbf{Proof}:
We proceed, from (\ref{AFLT}),
\bea\label{AGTbifund} &&_{Y'_1,Y'_2}\langle
P\ket{P}_{Y_1,Y_2}=\,\,
\!\!_{Y'_1,Y'_2}\bra{P}V_{\alpha=0}\ket{P}_{Y_1,Y_2}\\\nonumber\\\nonumber
&=& \prod_{Y_1}\{Q- [(a_{Y_1}+1)b^{-1}- l_{Y'_1}
b]\}\prod_{Y'_1}\{(a_{Y'_1}+1)b^{-1}- l_{Y_1}b\}\\\nonumber&\times&
\prod_{Y_1}\{Q- [2P+(a_{Y_1}+1)b^{-1}- l_{Y'_2}
b]\}\prod_{Y'_2}\{-2P+(a_{Y'_2}+1)b^{-1}-
l_{Y_1}b\}\\\nonumber&\times&\prod_{Y_2}\{Q- [-2P+(a_{Y_2}+1)b^{-1}-
l_{Y'_1} b]\}\prod_{Y'_1}\{2P+(a_{Y'_1}+1)b^{-1}-
l_{Y_2}b\}\\\nonumber&\times& \prod_{Y_2}\{Q- [(a_{Y_2}+1)b^{-1}-
l_{Y'_2} b]\}\prod_{Y'_2}\{(a_{Y'_2}+1)b^{-1}- l_{Y_2}b\}\,.\eea
We shall prove now that under this situation, if the result is non-zero, one can conclude\[\vec{Y} = \vec{Y'}.\] 
If
$$_{Y'_1,Y'_2}\langle  P\ket{P}_{Y_1,Y_2}\neq
0,$$ one gets $$ y_{1,1}\leq y'_{1,1}.$$\footnote{Here we use the
notation $y_{1,r}$ to label the $r$-th part of the partition $Y_1$.}
Since otherwise there must exist a box in the tableau $Y_1$
satisfying
$$a_{Y_1} = 0, \,\, l_{Y_1'}= - 1\, \, .$$ This will lead to
\be Q - [(a_{Y_1}+1)b^{-1}- l_{Y'_1}b] = 0,\ee
This
argument cycles and one finally conclude: $$ y_{1,i}\leq y'_{1,i}
\,\,, i=1,2,\dots\,. $$ For $Y_2$, similarly, the argument follows, and
gives:
$$y_{2,i}\leq y'_{2,i} \,\,, i=1,2,\dots \,. $$  However, the
original condition $$|Y'_1|+|Y'_2|=|Y_1|+|Y_2|$$ then forces $Y_1 = Y'_1,
Y_2 = Y'_2$. Q.E.D.

The orthogonality condition, (\ref{Bifund}), strongly suggests the existence of mutually commuting Hermitian operators, whose common eigenstates form a complete orthogonal basis of the Hilbert space. One of the operators, called the energy operator, probably cubic in $a_n$'s and $c_n$'s (since this is most likely the case beyond $I_2$), is the first object we are going to construct.
However, hermiticity alone is not enough to constrain the possible forms of the construction. For example, $H_0=H^+ + H^-$ is Hermitian, but does not belong to $Vir\oplus u(1)$. In fact, we shall show that altogether there should be at least 3 conditions $H$ are to meet in order the orthogonality of the $H_0$ eigenstates play an important role here.
  \beaa i)&&\hspace{1,5cm}Hermiticity\\
 ii)&&\hspace{1.5cm}Triangularity\\
iii)&&\hspace{1.5cm}Reflection-invariance\,.\eeaa

Now we explain what the other two conditions means. Condition ii), triangularity,
means that $H$, in its matrix form, $H_{\vec{Y'},\vec{Y}}(P)=\bra{\vec{Y'}}H(P)\ket{\vec{Y}}$, where $\ket{\vec{Y}}$'s are the eigenstates of $H_0= H^++H^-$(the ``unperturbed'' energy operator)\footnote{Here we have fixed $\hat{P}$ eigenvalue equals $P$ .}, is lower(or upper)-triangular with $H_I\equiv H-H_0$ strictly triangular (with zero diagonal entries). Under such circumstances, the spectrum of $H$ coincides with that of $H_0$,
and the eigenstates of $H$ can be expressed as $\ket{P}_{Y_1,Y_2} =\Omega_{Y_1,Y_2}(P)R(E)\ket{\vec{Y},P}$, here the normalization constant $\Omega_{Y_1,Y_2}(P)$ will be specified later on. $R(E)=1+\tilde{R}(E)$, a unitriangular matrix, is again triangular with identity diagonal entries following the triangularity of $H$.
$H^\pm$ is the collective mode Hamiltonian for the Calogero-Sutherland model in terms of the oscillators ${a^{\pm}_n}$'s. Thus the eigenstate of $H_0$ is just  $J^+_{-Y_1}J^-_{-Y_2}\ket{P}$.
$H^\pm$ in our formalism (including the zero modes $a^\pm_0$) is defined as
$$H^\pm = -i\frac{1}{3}\oint(z\partial_z\varphi^\pm(z))^3dz/z +\sum_{n=1}^{\infty}Qn a^\pm_{-n}a^\pm_n$$
Since $\varphi^\pm\dagger=\varphi^\mp$, we have $J^+_Y\dagger =J^-_{-Y}$.
There is a natural question on how to define the inner products between  $J^\pm_Y$'s.
The answer is that we need the condition iii) Reflection-invariance.
Notice that $\bra{P'}P\rangle$ not zero means $P+P'=0$. In order to get a non-vanishing result,
we need to shift $\bra{P}$ to $\bra{-P}$. We thus expect that there exists an operation which changes $_{Y_1,Y_2}\bra{P}$ to $_{Y_2,Y_1}\bra{-P}$. We call this operation reflection following the terminology in a similar situation in \cite{Belavin:2011js}. Actually, by looking closer to the NIC
 formula i.e. the r.h.s of (\ref{AFLT}), one can find that there exist an apparent symmetry
 \be_{Y_1,Y_2}\bra{P}\leftrightarrow _{Y_2,Y_1}\bra{-P}. \ee If we change either bra state  $\bra{P}_{Y_1,Y_2}$ to $_{Y_2,Y_1}\bra{-P}$,  or ket state $\ket{P}_{Y_1,Y_2}$ to $\ket{-P}_{Y_2,Y_1}$, on the l.h.s. of (\ref{AFLT}), the factors on the r.h.s. of eq(\ref{AFLT}) get  reshuffling but the final result keep invariant. We may
 name this symmetry ``reflection'' or ``flipping'' symmetry. On the 2d CFT side, from general reasoning that such an operation should be conformally invariant, it is natural to identify  the insertions of the screening charges as this ``reflection'' operation.
For Liouville theory (or Coulomb gas model),
we can attach to $V_{\alpha=0}$ some screening charges\footnote{We suppose originally there is no screening charge attached to $V_{\alpha=0}$ for simplicity.}, such that \bea
_{Y_1',Y_2'}\bra{P}V_{0}S^n\ket{P}_{Y_1,Y_2}\neq 0\\
S=\oint e^{2ib\varphi(z)}\d z \,.\eea Now the neutrality condition forces
\(2P+nb=0 \,\,\,.\)
If this is satisfied, then Felder's contour for the integration of the screening charges actually closes and $S^n$ becomes a floating charge\cite{Felder:1988zp,Bernard:1989iy}. Now $S^n$ can move away from $V_\alpha$ and communing through $L_n$'s and finally acts on the vacuum state $\bra{P}$. Since $S^n$ acts by not changing the  conformal weight, we deduce
$$\bra{P}S^n=\bra{-P}$$ for a suitable normalization of $S^n$.
Similar arguments apply to the case of $V_\alpha, \ \alpha\neq 0$, and one can always move a subset of screening charges, $S^{\frac{-2P}{b}}$ away from $V_\alpha(z)$.
Since AGT duality formula is valid for any $P$, we may assume that $n$ can
take arbitrary real value, as analytical continuation away from integer
$n$. This flipping is due to the fact that  $S^n$ can be detached
from $V_{\alpha}$, and act on the vacuum directly. Similar operation exists
in Felder's BRST cohomology \cite{Felder:1988zp}.

We are going to identify the reflection symmetry in NIC as the Hamiltonian symmetry in
2d CFT for the insertions of the screening charges $S^n$ with $2P = -n b$. Since Hamiltonian $H\in Vir\oplus u(1)$,
satisfies $\left[H,S^n\right]=0$ it has the property of double
degeneracy. So $S^n$ with $2P = -n b$ should  map one AFLT state to its partner
state. If we require
\bea_{Y_1,Y_2}\bra{P}H &=& \!\!
_{Y_1,Y_2}\bra{P}E_{Y_1,Y_2} \\
_{Y_1,Y_2}\bra{P}S^n H &=&
 _{Y_1,Y_2}\bra{P}S^n E_{Y_1,Y_2}\,, \eea
then  we can identify
$$ _{Y_1,Y_2}\bra{P} S^n = _{Y_2,Y_1}\bra{-P}\,,$$ since reflection symmetry means $E_{Y_1,Y_2}(P)=E_{Y_2,Y_1}(-P)$.
Notice that  nothing has changed for the $u(1)$ part. Define $P^\pm =-i a_0 \mp i c_0, $ then we have\footnote{We have set $a_0\ket{P} = 0$ throughout this paper.}:
\[P^\pm\ket{P} = \pm P\ket{P},\,\,\, \bra{-P}P^\pm = \bra{-P}(\pm P)\,, \] which
obviously shows that $\langle -P\ket{P}\neq 0$. So reflection invariance means that we can identify the inner product $_{Y_1',Y_2'}\bra{P}P\rangle_{Y_1,Y_2}$ with either $_{Y_2',Y_1'}\bra{-P}P\rangle_{Y_1,Y_2}$ or $_{Y_1',Y_2'}\bra{P}-P\rangle_{Y_2,Y_1}$ by the incertions of screening charges satisfying $n=-2Pb^{-1}$.

Having determined that $\ket{P}_{\vec{Y}}$ form a normalizable orthogonal basis, the next step is the determination of their normalization. Before doing this, let's review the so-called Carlsson-Okounkov formula\cite{Carlsson-Okounkov} which is useful for our formulation.  First, define \bea E= 1+e_1 +e_2+\dots =
e^{-\sum_n\frac{(-)^n}{n}p_n} = e^{-\frac{1}{k}\varphi_{(-)}(-1)} \eea
\footnote{For infinitely many arguments $z_i$'s, $i=1,2,\cdots, \infty$, one may identify $p_n\equiv\sum_i z_i^n$ with $\frac{a_{-n}}{k}$, $k^2=\beta$ and $J_Y^{1/\beta}(\{p_n\})$ with $J_Y^{1/\beta}(\{\frac{a_{-n}}{k}\})$. Here our convention is that $\frac{a_{-n}}{k}\ket{0}$ creates a state  $\ket{p_n}$.
As a consequence, $e_n$ is to be identified with $P_{-1^n}\equiv \dfrac{J_{1^n}^{1/\beta}(\{\frac{a_{-n}}{k}\})}{n!}\equiv J_{-1^n}/n!$ . Such kind of  identification is justified because they share the same values of their inner products. For more details see \cite{Wu:2011cy}.}which is a vertex operator, and also a generating function for
$J_{-1^n}$\[e^{-\frac{1}{k}\varphi_{(-)}(z)}\ket{0} = \sum_n
(-)^n\dfrac{J_{-1^n}}{n!}z^n\ket{0}\,,\] here $e_i$ are elementary
symmetric functions, $p_n$ is the power sum symmetric function. Then
\be e^{-\frac{1}{k}\varphi_{(-)}(-1)}\ket{0} =
\sum_n\dfrac{J_{-1^n}}{n!}\ket{0} \equiv \sum_n
P_{-1^n}\ket{0}. \ee The conjugation of $E$ reads \be E^{\ast}
=e^{\frac{1}{k}\varphi_{(+)}(-1)}\,,\ee and we have $$\bra{0} E^{\ast} =
\bra{0}\sum_n P_{1^n}\,.$$ Now the Carlsson-Okounkov formula reads
\bea &&\langle  E^m (E^{\ast})^{\beta-m-1} J_{-Y_1}, J_{-Y_2}\rangle
\\\nonumber
&=&(-)^{|Y_1|}\beta^{-|Y_1|-|Y_2|}\prod_{Y_1}(m+(a_{Y_1}+1)+\beta
l_{Y_2})\prod_{Y_2}(m-a_{Y_2}-\beta(l_{Y_1}+1)) \\\nonumber&=&
\langle  J_{Y_1} E^{\beta-m-1}(E^{\ast})^m J_{-Y_2}\rangle
\\\nonumber&=&\langle
J_{Y_1}e^{(-k+k^{-1}+\frac{m}{k})\varphi_{(-)}(-1)}e^{\frac{m}{k}\varphi_{(+)}(-1)}J_{-Y_2}\rangle
\,. \eea For Liouville theory, $k=-ib$. If we set $\frac{m}{-ib}=-2i\alpha$, then the Carlsson-Okounkov
formula reads \bea &&\langle
J_{Y_1}e^{i(Q-2\alpha)\varphi_{(-)}(-1)}e^{-2i\alpha\varphi_{(+)}(-1)}J_{-Y_2}\rangle
\\\nonumber&=&
(-)^{|Y_2|}b^{-|Y_1|-|Y_2|}\prod_{Y_1}(-2\alpha+(a_{Y_1}+1)b^{-1}-l_{-Y_2}b)
\prod_{Y_2}(-2\alpha-a_{Y_2}b^{-1}+(l_{Y_1}+1)b)\,.\eea

The normalization of the AFLT states is inherited from AFLT's version of the AGT duality formula, (\ref{AFLT}) and the orthogonality condition, (\ref{AGTbifund}),
\bea\label{norm1} && _{Y_1,Y_2}\langle
P\ket{P}_{Y_1,Y_2}= _{Y_2,Y_1}\langle
-P\ket{P}_{Y_1,Y_2}\\\nonumber
&=&\prod_{Y_1}\{-a_{Y_1}b^{-1}+(l_{Y_1}+1)b\}\{(a_{Y_1}+1)b^{-1}-l_{Y_1}b\}\\\nonumber
&\times&\prod_{Y_2}\{-a_{Y_2}b^{-1}+(l_{Y_2}+1)b\}\{(a_{Y_2}+1)b^{-1}-l_{Y_2}b\}\\\nonumber
&\times&\prod_{Y_1}\{-2P-a_{Y_1}b^{-1}+(l_{Y_2}+1)b)\}\prod_{Y_2}\{-2P+(a_{Y_2}+1)b^{-1}-l_{Y_1}b\}\\\nonumber
&\times&\prod_{Y_2}\{2P-a_{Y_2}b^{-1}+(l_{Y_1}+1)b)\}\prod_{Y_1}\{2P+(a_{Y_1}+1)b^{-1}-l_{Y_2}b\}\\\nonumber
&=&(-)^{|Y_1|+|Y_2|}j_{Y_1}j_{Y_2}\\\nonumber&\times&
\prod_{Y_1}\{-2Pb-
a_{Y_1}+(l_{Y_2}+1)b^2\}\prod_{Y_2}\{-2Pb+(a_{Y_2}+1)-l_{Y_1}b^2\}\\\nonumber&
\times&\prod_{Y_2}\{2Pb-a_{Y_2}+(l_{Y_1}+1)b^2\}\prod_{Y_1}
\{2Pb+(a_{Y_1}+1)-l_{Y_2}b^2\}\\\nonumber
&\equiv& j_{Y_1}j_{Y_2}\Omega_{Y_2, Y_1}(-P) \Omega_{Y_1, Y_2}(P)\\\nonumber
&=&j_{Y_1}j_{Y_2}
\langle  J_{Y_2} e^{i(Q-2P)\varphi_{(-)}(-1)}e^{-i2P\varphi_{(+)}(-1)}
J_{-Y_1}\rangle
\\\nonumber&\times&(b^4)^{|Y_1|+|Y_2|}\langle  J_{Y_1} e^{i(Q+2P)\varphi_{(-)}(-1)}e^{i2P\varphi_{(+)}(-1)}
J_{-Y_2}\rangle \,. \eea
In reaching the last line in the above eqation, Carlsson-Okounkov formula has been applied, and we have defined
\bea\label{norm2}
\Omega_{Y_1, Y_2}(P) \nonumber
&=&
(-)^{|Y_1|}b^{|Y_1|+|Y_2|}\prod_{Y_1}\left(2P+(a_{Y_1}+1)b^{-1}-l_{Y_2}b\right)
\prod_{Y_2}\left(2P-a_{Y_2}b^{-1} +(l_{Y_1}+1)b\right)
\\\nonumber&=&(-b^2)^{(|Y_1|+|Y_2|)}\langle
J_{Y_1}e^{i(Q+2P)\varphi_{(-)}(-1)}e^{i2P\varphi_{(+)}(-1)} J_{-Y_2}\rangle \\
&=&b^{2(|Y_1|+|Y_2|)}\langle
J_{Y_1}e^{i(Q+2P)\varphi_{(-)}(1)}e^{i2P\varphi_{(+)}(1)} J_{-Y_2}\rangle \\\nonumber
&\equiv&\Omega_{\vec{Y}}(P)\, .
\eea
Notice that $\Omega_{\vec{Y}}(P)$ is just a generalization of $\Omega_{Y}(P)$ defined in \cite{AFLT}.
\section{The Construction of the AFLT States}\
Now we come to our main problem of the construction of the Hamiltonian $H$ with
the requirement that its eigenstates be identified with ALFT states satisfying
(\ref{AFLT}). We prefer to work first on the basis of Jack symmetric functions $J_{\vec{Y}}$,
which already form an orthogonal basis. We found that if $H_I$ matrix elements are strictly triangular on this basis, then the orthogonality of the $H=H_0+H_I$ eigenstates follows immediately
from the orthogonality of the $H_0$ eigenstates. This is just the simplest way to go from  one orthogonal basis to another one. To see this, let's introduce an operator $R(E)$ which map $H_0$ eigenstates to $H$ eigenstates, with $\Omega_{Y_1,Y_2}(P)$ the normalization constant
\bea\label{Roperator}
\ket{P}_{Y_1,Y_2} &=& R(E) J_{-Y_1}^+
J_{-Y_2}^-\ket{P}\Omega_{Y_1,Y_2}(P)\,\\\nonumber
_{Y_2',Y_1'}\bra{-P} &=& \bra{-P} J_{Y_2'}^- J_{Y_1'}^+R(E')^{\dagger}
\Omega_{Y_2'Y_1'}(-P)\\\nonumber R(E)&=&1+\cdots = 1+\tilde{R}(E)\,,
\eea
where the reflection symmetry has been applied to the AFLT states
$$_{Y_1',Y_2'}\bra{P}S^n  = \bra{P} S^n J_{Y_1'}^- J_{Y_2'}^+R(E')^{\dagger}
\Omega_{Y_1'Y_2'}(P)= \\
_{Y_2',Y_1'}\bra{-P} = \bra{-P} J_{Y_2'}^- J_{Y_1'}^+R(E')^{\dagger}
\Omega_{Y_2'Y_1'}(-P) ,$$  and $\tilde{R}(E)$ is strictly lower(or upper)-triangular $\Rightarrow$ $\tilde{R}_{\vec{Y},\vec{Y}}(E)=0$. The way $R(E)$ is expanded in (\ref{Roperator}) follows
from the normalization condition, (\ref{norm1}-\ref{norm2}) as we shall see in (\ref{norm3}).

The Hermitian operator $H$ should satisfy:
\bea H\ket{P}_{Y_1,Y_2} &=&
E_{Y_1,Y_2}(P)\ket{P}_{Y_1,Y_2} \\\nonumber _{Y_2',Y_1'}\bra{-P}H &=&
_{Y_2',Y_1'}\bra{-P}E_{Y_2', Y_1'}(-P), \eea where the energy
eigenvalue has the double degeneracy: \be E_{Y_2, Y_1}(-P) = E_{Y_1,Y_2}(P)
\,,\ee due to the
orthogonality condition, \bea\label{norm3} _{Y_1',Y_2'}\langle  P\ket{P}_{Y_1, Y_2} = _{Y_2', Y_1'}\bra{-P}P\rangle _{Y_1, Y_2}\propto\delta_{\vec{Y},\vec{Y'}}\,.\eea Then the next step is to
determine if we get the right normalization for $\ket{P}_{Y_1,Y_2}$ \bea\label{norm3}
_{Y_2,Y_1}\langle {-P}\ket{P}_{Y_1,Y_2}
&=&\bra{-P}J_{Y_2}^-J_{Y_2}^+ R(E)^{\dagger} R(E) J_{-Y_1}^+
J_{-Y_2}^-\ket{P} \Omega_{Y_1, Y_2}(P)\Omega_{Y_2, Y_1}(-P) \\\nonumber  &=& \bra{-P} J_{Y_2}^-J_{Y_1}^+(1+\tilde{R}(E)^{\dagger})(1+\tilde{R}(E))J_{-Y_1}^+J_{-Y_2}^-\ket{P}\Omega_{Y_1, Y_2}(P)\Omega_{Y_2, Y_1}(-P)\\\nonumber  &=&
\Omega_{Y_1, Y_2}(P)\Omega_{Y_2, Y_1}(-P) \\\nonumber 
&\times&\left[j_{Y_1}j_{Y_2}+
\bra{-P}J_{Y_2}^-J_{Y_1}^+(\tilde{R}(E)^\dagger + \tilde{R}(E)
+\tilde{R}(E)^\dagger \tilde{R}(E))J_{-Y_1}^+J_{-Y_2}^-\ket{P}\right] \\\nonumber 
&=& \Omega_{Y_1, Y_2}(P)\Omega_{Y_2, Y_1}(-P)j_{Y_1}j_{Y_2}\,.\eea
It is in agreement with (\ref{norm1}). In deriving this we have used the fact 
that if $R(E)$ is a unitriangular matrix\footnote{A unitriangular matrix is a triangular matrix with  the diagonal entries equal to 1 .}, \be\label{Raction}
R(E)J_{-Y_1}^+J_{-Y_2}^-\ket{P} = J_{-Y_1}^+J_{-Y_2}^-\ket{P}+
\sum_{\begin{subarray}{c}|Y_1'|>|Y_1|\\|Y_2'|<|Y_2|\\|Y_1'|+|Y_2'|=|Y_1|+|Y_2|\end{subarray}}
R_{Y_1, Y_2}^{Y_1' Y_2'}(E)J_{-Y_1'}^+J_{-Y_2'}^-\ket{P}\,,\ee then it is
easy to check that \beaa
&&\bra{-P}J_{Y_2}^-J_{Y_1}^+\tilde{R}(E)^{\dagger}J_{-Y_1}^+J_{-Y_2}^-\ket{P}
\\&&=\bra{-P}J_{Y_2}^-J_{Y_1}^+\tilde{R}(E)J_{-Y_1}^+J_{-Y_2}^-\ket{P}\\&&=
\bra{-P}J_{Y_2}^-J_{Y_1}^+\tilde{R}(E)^{\dagger}\tilde{R}(E)J_{-Y_1}^+J_{-Y_2}^-\ket{P}\\&&=0\,.
\eeaa
Now we summarize the requirements for $R(E)$
\begin{eqnarray*}
&&i)\hspace{1.0cm} \text{$R(E)$ is unitriangular} \\
&&ii)\hspace{1.0cm} \text{$R(E)$ creates the eigenstate for $H$}\\
&& \hspace{1.0cm}HR(E)J_{-Y_1}^+J_{-Y_2}^-\ket{P} = E_{Y_1, Y_2}(P)
R(E)J_{-Y_1}^+J_{-Y_2}^-\ket{P}\\\
&&iii)\hspace{1,0cm} \text{Reflection invariant}\\ &&\hspace{1.0cm} S^n
R(E) J_{-Y_1}^+ J_{-Y_2}^-\ket{P}\Omega_{Y_1, Y_2}(P)= R(E)
J_{-Y_2}^+J_{-Y_1}^-\ket{-P}\Omega_{Y_2,Y_1}(-P)\\&&\hspace{1.0cm}
\left[S^n, H\right] = 0, \ \ \ \  E_{Y_1, Y_2}(P)= E_{Y_2, Y_1}(-P)
\end{eqnarray*}
This means that the Hamiltonian $H$ should also be triangular, and $H_I$ strictly triangular,
\bea H &=& H^+ +
H^- + H_I\\
H^{\pm}&=&\dfrac{-i}{3}\oint\left(z\partial_z\varphi^{\pm}\right)^3
+ \sum_{n>0}Q n a_{-n}^{\pm}a_n^{\pm}\,,\eea here $H_I$ is to be
determined later on. \beaa H^{+} +H^-&=& \dfrac{-i}{3}\oint
\left[\left(z\partial_z(\varphi+\tilde{\varphi})\right)^3+\left(z\partial_z(\varphi-\tilde{\varphi}) \right)^3\right]
\dfrac{\d z}{z}+\\&&\sum_{n>0}Q n
a_{-n}^{+}a_n^{+}+\sum_{n>0}Q n
a_{-n}^{-}a_n^{-}\\&=&\dfrac{-i}{3}\oint\left[2\left(z\partial_z\tilde{\varphi}\right)^3
+ 6(z\partial_z\tilde{\varphi})(z\partial_z\varphi)^2\right]\dfrac{\d
z}{z}
\\&+& \sum_{n>0}2Q n(a_{-n}a_n+ c_{-n}c_n) \\&=& \dfrac{-2i}{3}\oint\left(z\partial_z\tilde{\varphi}\right)^3 \frac{\d z}{z}  +
\sum_{n>0}2Q n a_{-n}a_n - 2i \oint
\left(z\partial_z\tilde{\varphi}\right)\left(z\partial_z\varphi\right)^2\frac{\d z}{z} + 2\sum_{n>0} Q n
c_{-n}c_n\,.\eeaa

Now the requirement that $H$ commute with $S^n$ is equivalent to say
that $H$ can be written in terms of $L_n$'s and $a_n$'s.
To make $H$ triangular, we may try \beaa H_I&\propto& \sum_n Q n a_{-n}^+
a_n^-\\&=& \sum_n Q n (a_{-n}a_n-c_{-n}c_n - a_{-n}c_n +c_{-n}a_n)
\\&=& \sum_n Q n (a_{-n}a_n-c_{-n}c_n)+Q\oint
z\partial_z\tilde{\varphi}(z\partial_z)^2\varphi \frac{\d z}{z}\, . \eeaa If we now make use of (\ref{TZ1}) and choose
\[H_I = \sum_n 2Q n a_{-n}^+a_n^-,\,\] 
then we get
\bea 
H&=& -\dfrac{2i}{3}\oint(z\partial_z\tilde{\varphi})^3\dfrac{\d
z}{z} + 4Q \sum_{n\in\mathbb{N}^+}n a_{-n} a_n -
2i\oint(z\partial_z\tilde{\varphi}) z^2 T(z)\dfrac{\d
z}{z} +2ia_0\dfrac{Q^2}{4}
\\\nonumber
&=& -\dfrac{2i}{3}\oint(z\partial_z\tilde{\varphi})^3\dfrac{\d
z}{z} + 4Q \sum_{n\in\mathbb{N}^+}n a_{-n} a_n -
2i\sum_{n\in\mathbb{Z}}a_{-n}L_{n}+2ia_0\dfrac{Q^2}{4}
\\\nonumber&=&-i\left\{\sum_{n,m\in\mathbb{N}^+}\left(a_{-n-m}^+ a_n^+ a_m^+
+a_{-n-m}^-a_n^-a_m^-\right)\right.
\\\nonumber&+&\left.\sum_{n,m\in\mathbb{N}^+}\left(a_{-n}^+ a_{-m}^+ a_{n+m}^+
+a_{-n}^-a_{-n}^-a_{n+m}^-\right)\right\}\\&+&
\sum_{n\in\mathbb{N}^+} Q n (a_{-n}^+a_n^++a_{-n}^-a_n^-
+2a_{-n}^+a_n^-)\\&+&\sum_{n\in\mathbb{N}^+}-2i a_0^+a_{-n}^+a_n^+ -
2i a_0^- a_{-n}^- a_n^- -\dfrac{i}{3}((a_0^+)^3+(a_0^-)^3) \,.\eea
Clearly, $H$ indeed satisfies the three requirements proposed in the previous section.
\beaa i)&&\hspace{1,5cm}Hermitian\\
 ii)&&\hspace{1.5cm}Triangular\\
iii)&&\hspace{1.5cm}Reflection-invarint\,.\eeaa

Besides, we found that $H\propto I_3$, where $I_3$ is defined in
Appendix C of \cite{AFLT} as one of the
infinitely many commuting operators which may makes the system
integrable. The authors of \cite{AFLT} have checked for the first a few levels
that the AFLT states,
$\ket{P}_{Y_1,Y_2}$, which satisfies AGT duality formula, (\ref{AFLT}), are also the
eigenstates of $I_3$. But a general formula for the $I_3$ eigenstates is missing in \cite{AFLT}.

Now the next question is : how to find all the eigenstates of $H$?
First, let's consider $H^+$ \bea H^+ &=& -i \sum_{n,m\in\mathbb{N}^+} \left\{a_{-n-m}^+
a_n^+ a_{m}^+ + a_{-n}^+a_{-m}^+
a_{n+m}^+\right\}\\\nonumber&+&\sum_{n\in\mathbb{N}^+} \left\{ n Q a_{-n}^+ a_n^+ + 2a^+_0(-i)
a_{-n}^+a_n^+\right\} - \frac{i (a_0^+)^3}{3}\,,\eea
Its eigenvalue
$$H^{+}J^{+}_{-Y}|P^{+}\rangle =E^{+}_{Y}(P^{+})J^{+}_{-Y}|P^{+}\rangle $$
$$E^{+}_{Y}(P^+)=\sum_i\left\{y^{2}_{i}b^{-1}+(2i-1)y_i b\right\}+2P^{+}|Y|-\frac{(P^{+})^3}{3}.$$
Here we have assumed the zero modes take the following eigenvalues,
\be
a_0=iP^a\, ,\, c_0=iP^c \, ,\, a^\pm_0=iP^\pm=i(P^a\pm P^c)
\ee
For the bi-Jack system, we have the following eigenequation,
$$HR(E)J^{+}_{-Y_1}J^{-}_{-Y_2}|P^{+},P^{-}\rangle =E_{Y_1,Y_2}(P^{+},P^{-})R(E)J^{+}_{-Y_1}J^{-}_{-Y_2}|P^{+},P^{-}\rangle\,. $$
Triangularity means
\begin{eqnarray*}
E_{Y_1,Y_2}(P^{+},P^{-})=&&E^{+}_{Y_1}(P^{+})+E^{-}_{Y_2}(P^{-})\\
                        =&&\sum_{i} \left\{y_{1,i}^2 b^{-1}+(2i-1)y_{1,i}^2 b\right\} +\sum_{i}\left\{y_{2,i}^2 b^{-1}+(2i-1)y_{2,i}^2 b\right\} \\
                         &&+2P^{+}|Y_1|+2P^{-}|Y_2|-\frac{(P^{+})^3+(P^{-})^3}{3}
\end{eqnarray*}
Since $H$ can be constructed in terms of $L_n$'s and $a_n$'s, so
$S^n|P^{+},P^{-}\rangle _{Y_1,Y_2}$ dose not change
the eigenvalue. But $S^n$ changes $P^c\rightarrow-P^c$ and
$P^{+}\leftrightarrow P^{-}$ and from
$E_{Y_1,Y_2}(P^{+},P^{-})=E_{Y_2,Y_1}(P^{-},P^{+})$. We conclude
\be\label{Sn}
S^n|P^{+},P^{-}\rangle _{Y_1,Y_2}\propto|P^{-},P^{+}\rangle _{Y_2,Y_1}.
\ee
Next, since $P^a$ does not play any important role, we may consider
it as a gauge symmetry and can be fixed to any desired value. For
convenience, we fix $P^{a}=0$, hence, $P^+=P^c\equiv P$, $
P^{-}=-P^c\equiv-P $ and
\begin{eqnarray*}
E_{Y_1,Y_2}(P,-P)&&\equiv E_{Y_1,Y_2}(P)\\
                 &&=E_{Y_1}+E_{Y_2}+2P(|Y_1|-|Y_2|)
\end{eqnarray*}
Here
$$E_Y=\sum_i \left\{y^{2}_{i}b^{-1}+(2i-1)y_i b\right\}$$
If we define
\begin{eqnarray*}
          |P\rangle  &\equiv&|P,-P\rangle  \\
|P\rangle _{Y_1,Y_2} &\equiv&R(E)J^{+}_{-Y_1}{J^{-}_{-Y_2}}|P\rangle
\Omega_{Y_1,Y_2}(P)
\end{eqnarray*}
Then we infer from from (\ref{Sn})
$$S^n|P\rangle _{Y_1,Y_2}=|-P\rangle _{Y_2,Y_1}$$ with the proper normalization for $S^n$.
Now we are going to determine $R(E)$ which satisfies
$$HR(E)J^{+}_{-Y_1}{J^{-}_{-Y_2}}|P\rangle =E_{Y_1,Y_2}(P)R(E)J^{+}_{-Y_1}{J^{-}_{-Y_2}}|P\rangle .$$
\begin{propn}
$$R(E)J^{+}_{-Y_1}{J^{-}_{-Y_2}}|P\rangle =\frac{1}{1-\frac{1}{E-H_0}H_I}J^{+}_{-Y_1}{J^{-}_{-Y_2}}|P\rangle ,$$
here $H_0=H^{+}+H^{-},E=E_{Y_1,Y_2}(P)$.
$R(E)$ defined in such a way should be understood as
\begin{eqnarray*}
  R(E)&=&\frac{1}{1-\frac{1}{E-H_0}H_I} \\
             &=& \sum_{n=0}^{\infty}(\frac{1}{E_{Y_1,Y_2}(P)-H_0}H_I)^n
\end{eqnarray*}
\end{propn}

\hspace{-0.6cm}\textbf{Proof}:\ \ \ \ First, we rewrite $H$ as
\begin{eqnarray*}
 H&=&H_0+H_I \\
  &=& E+H_0+H_I-E \\
&=& E+(H_0-E)(1+\frac{1}{H_0-E}H_I)\,.
\end{eqnarray*}
Then from
\begin{eqnarray*}
 HR(E)
&=& (E+(H_0-E)(1+\frac{1}{H_0-E}H_I))\frac{1}{1-\frac{1}{E-H_0}H_I}\\
&=& E\frac{1}{1-\frac{1}{E-H_0}H_I}+H_0-E\,,\\
&=& ER(E)+H_0-E\, ,
\end{eqnarray*}
one gets
$$HR(E)J^{+}_{-Y_1}{J^{-}_{-Y_2}}|P\rangle =
 ER(E)J^{+}_{-Y_1}{J^{-}_{-Y_2}}|P\rangle +(H_0-E)J^{+}_{-Y_1}{J^{-}_{-Y_2}}|P\rangle \,.$$
Since $J^{+}_{-Y_1}{J^{-}_{-Y_2}}|P\rangle $ is an eigenstate of $H_0$ with eigenvalue
$E=E_{Y_1,Y_2}(P)$, we have
$$(H_0-E)J^{+}_{-Y_1}{J^{-}_{-Y_2}}|P\rangle =0.$$
Hence, we conclude that $R(E)J_{-Y_1}^+J_{-Y_2}^-\ket{P}$ is an eigenstate of $H$ with eigenvalue $E$,
$$ E \equiv E_{Y_1,Y_2}(P)=\left(\sum_{i=1,l=1}^{i=2,l=y_{i,1}^t}(y_{i,l})^2+(2i-1)y_{i,l}\right) +2P(|Y_1|-|Y_2|)\,,$$ Q.E.D.\\
Now we shall address the question raised in \cite{AFLT} on the possible degeneracy of $H$. The authors of \cite{AFLT}, argued that
$I_3$ has some degeneracy at level 4 and higher. We have analyzed
what causes such kind of degeneracy. After analyzing the spectrum of
$H$, we believe that such degeneracy  happens when $|Y_1|=|Y_2|$,
and we have $2P(|Y_1|-|Y_2|) = 0$,
$$ E_{Y_1,Y_2}(P)=E_{Y_1}+E_{Y_2}=E_{Y_2}+E_{Y_1}=E_{Y_2,Y_1}(P)$$
This can happen, for $Y_1\neq Y_2$, first at level 4,
$|Y_1|+|Y_2|\equiv |\vec{Y}|=4$, and $Y_1=2$, $Y_2=1^2$. Such
degeneracy can happen at any even level higher or equal to 4. For example
at $\textrm{level}=6$:
$$Y_1=3, Y_2=1^3, \textrm{or}~ Y_1=3, Y_2=\{2,1\}, \textrm{or}~ Y_1=1^3, Y_2=\{2,1\},$$
or simply, we have $(Y_1,Y_2)$ pair
$$(3,1^3),(3,\{2,1\}),(1^3,\{2,1\})$$
Such degeneracy does not cause any problem in constructing the
eigenstate of $H$ for the following reasons.

i) The mother state $J^{+}_{-Y_1}J^{-}_{-Y_2}|P\rangle $ is
uniquely determined by the Young diagram, even for the degenerate
$E$.

ii) Consider power expansion
$$R(E)=\sum_{n=0}^{\infty}(\frac{1}{E_{Y_1,Y_2}(P)-H_0}H_I)^n.$$
For an intermediate state.
\begin{eqnarray*}
E_{Y_1,Y_2}(P)-H_0&&\sim E_{Y_1,Y_2}(P)-E_{Y^{'}_1,Y^{'}_2}(P)\\
                  &&=E_{Y_1}+E_{Y_2}-E_{Y^{'}_1}-E_{Y^{'}_2}+2P(|Y_1|-|Y_2|-|Y^{'}_1|+|Y^{'}_2|)
\end{eqnarray*}
Since $|Y^{'}_1|>|Y_1|$, $|Y^{'}_2|<|Y_2|$ and
$|Y_1|-|Y^{'}_1|+|Y^{'}_2|-|Y_2|<0$ because of strictly triangularity of $H_I$, so for a general value of $P$,
$\frac{1}{E_{Y_1,Y_2}(P)-H_0}$ is not singular, and
$R(E)J^{+}_{-Y_1}{J^{-}_{-Y_2}}|P\rangle $ is well defined.

iii) The construction given above leads to the orthogonality of the
state $|P\rangle _{Y_1,Y_2}$ for distinct $Y_1,Y_2$ even for the
degenerate values of $E$, cf. eqs.(\ref{norm1},\ref{norm2},\ref{norm3}).

iv) It can be proven that the eigenstate of $H$, constructed as in proposition 2,
is actually the common eigenstate for all the conseved charges which commute with
$H$, with the mild assumption that all the conserved charges are triangular in a similar way as $H$ is. Due to lack of space for the present paper, we shall give a proof on this statement elsewhere.

Finally, we shall make a comment on the possible poles of $R(E)$ in the complex $p-$plane.  
The $R(E)$ matrix elements is calculated based on the following formula,
\begin{eqnarray*}
  |P\rangle _{Y_1,Y_2} &&= R(E)J^{+}_{-Y_1}{J^{-}_{-Y_2}}|P\rangle \Omega_{Y_1,Y_2}(P)  \\
  &&=\sum_{n=0}^{\infty}(\frac{1}{E_{Y_1,Y_2}(P)-H_0}H_I)^nJ^{+}_{-Y_1}{J^{-}_{-Y_2}}|P\rangle\Omega_{Y_1,Y_2}(P)\,.
\end{eqnarray*}
which always ends up with finite order perturbation because $H_I$ is strictly triangular.
We found, by the explicit calculations carried out so far, that
there is no pole in the finite $P$ complex plane. The poles in $R(E)$
either cancels the zeros in $\Omega_{Y_1,Y_2}(P) $ or simply cancels by summing over all the relevant terms. Of course, this property is also the necessary condition if $\ket{P}_{Y_1,Y_2}$'s satisfy (\ref{AFLT}).
 Now the
general AFLT state can be written as
\bea\label{Jackconstruction}\ket{P}_{Y_1, Y_2} &=&
\left\{\Omega_{Y_1,Y_2}(P)J_{-Y_1}^+J_{-Y_2}^- + \sum_{\begin{subarray}||Y_1'|= |Y_1|+1\\|Y_2'| = |Y_2|-1\end{subarray}}C_{Y_1,Y_2}^{Y_1',Y_2'} J_{-Y_1'}^+J_{-Y_2'}^-
\right.\\\nonumber&+& \sum_{\begin{subarray} ||Y_1''|= |Y_1|+2\\|Y_2''| = |Y_2|-2\end{subarray}}C_{Y_1,Y_2}^{Y_1'',Y_2''}
J_{-Y_1''}^+J_{-Y_2''}^-\\\nonumber&+&\left.\cdots+\sum_{|Y|=|Y_1|+|Y_2|}C_{Y_1,Y_2}^{Y,\varnothing}J_{-Y}^+\right\}\ket{P},\eea here
$C_{Y_1,Y_2}^{Y_3,Y_4}$ is the transition coefficient which measures
the changing from the Young tableau vector $(Y_1,Y_2)$ to
$(Y_3,Y_4)$.

We have calculated those coefficients up to level 4, the explicit results(up to level 3) are included in Appdix A.   
With the coefficients we calculated, one can
check that: \bea\label{CheckAGT} &&Z_{bif}(\alpha|P',\vec{X}; P,
\vec{Y}) =
\\\nonumber && \sum_{(X_1',X_2'),(Y_1',Y_2')} \bra{P'}
J^-_{X_1'}J^+_{X_2'} C_{X_1,X_2}^{X_1',X_2'}V_\alpha C_{Y_1,Y_2}^{Y_1',Y_2'}
J_{-Y_1'}J_{-Y_2'}\ket{P}\,,\eea holds true, thus (\ref{AFLT}) is verified. Here for simplicity,
we have only verified the cases without the incertions of the screening charges, i.e. $P+P'+\alpha=0$.

\section{Conclusion and Perspective}

The present work can be generalized in different ways. First, since the one parameter Jack symmetric function is a special limit of the two parameter Macdonald symmetric function, we expect that much of our work can be generalized to the cases where Macdonald symmetric function plays a role. In that case, we expect a similar relation to the NIC for 5d theory.  Second, the Calogero-Sutherland model is an integrable system. And consequently, Jack symmetric function is the common eigenstate of the infinitely many
commuting charges which are deformed $W^\infty$ charges. And for the construction of the AFLT states, the conserved charges are further deformed from those for the Jack symmetric functions. The final construction should give the same results as
$I_n$ proposed in \cite{AFLT}, which are constructed from integrable KdV equations. We find in this case, the AFLT states remain to be the eigenstates for all the conserved charges. However, it is desirable to have infinitely many conserved charges constructed explicitly.
Third, the reflection symmetry studied in this paper is actually powerful enough to give a closed form for the construction of the AFLT states. We shall present
this result in our future work. Another interesting idea related to our work is to consider the Jack function as a perturbation away from the Schur function, we have found that similar formalism applies \cite{Wu:2011dz}.
Finally, it is very interesting to see how  we present the full pants diagram for the conformal blocks, comparing to the one we have considered
with one external leg.

\section{Acknowledgments}
We are grateful to Yi-hong Gao for detailed explanations on the the AGT conjecture and the related topics.\\
This work is part of the CAS program ``Frontier Topics in Mathematical Physics'' (KJCX3-SYW-S03) and
is supported partially by a national grant  NSFC(11035008).

\appendix
\section{Coefficients for AFLT States(up to level 3)}
Now we give the explicit construction of the AFLT states up to
level 3. The transition coefficients $C_{Y_1,Y_2}^{Y_1',Y_2'}$ are defined as, 
$$C_{Y_1,Y_2}^{Y_1',Y_2'}\equiv
R_{Y_1,Y_2}^{Y_1',Y_2'}(E)\Omega_{Y_1,Y_2}(P),\ \ \ \ C_{Y_1,Y_2}^{Y_1,Y_2}\equiv \Omega_{Y_1,Y_2}(P)\,.$$

Level 2 coefficients:\\

\(\hspace{-0.6cm}C_{0,1^2}^{1,1}=C_{0,2}^{1,1}=-4 b \left(1+b^2\right) P,\\
C_{1,1}^{1^2,0}=\frac{\left(1+b^2\right) (1+2 b P)}{-1+b^2},\\
C_{1,1}^{2,0}=-\frac{b^2 \left(1+b^2\right) (1+2 b P)}{-1+b^2},\\
C_{0,1^2}^{12,0}=1+b^2 \left(3+2 b \left(b-\frac{2 \left(1+b^2\right) P}{-1+b^2}\right)\right),\\
C_{0,1^2}^{2,0}=\frac{4 b^3 \left(1+b^2\right) P}{-1+b^2},\\
C_{0,2}^{2,0}=\frac{\left(1+b^2\right) P^2 \left(-2+b^2+b^4+4 b P\right)}{-1+b^2},\\
C_{0,2}^{1^2,0}=\frac{4 b \left(1+b^2\right) P}{-1+b^2}\)\\

Level 3 coefficients:\\

\(\hspace{-0.6cm} C_{2,1}^{3,0}=\left\{-\frac{b^3 \left(1+b^2\right) (b+2 P) \left(1+b^2+2 b P\right)}{-2+b^2}\right\},\\
C_{2,1}^{\{2,1\},0}=\left\{\frac{4 \left(1+b^2\right) (1+b P) \left(1+b^2+2 b P\right)}{-2+b^2}\right\},\\
C_{1^2,1}^{1^3,0}=\left\{\frac{\left(1+b^2\right) (1+2 b P) \left(1+b^2+2 b P\right)}{-1+2 b^2}\right\},\\
C_{1^2,1}^{\{2,1\},0}=\left\{-\frac{4 b^3 \left(1+b^2\right) (b+P) \left(1+b^2+2 b P\right)}{-1+2 b^2}\right\},\\
C_{1,2}^{2,1}=\left\{-\frac{2 b^2 \left(1+b^2\right) (1+2 b P) \left(-1+b^2+2 b P\right)}{-1+b^2}\right\},\\
C_{1,2}^{1^2,1}=\left\{\frac{4 b \left(1+b^2\right) P (1+2 b P)}{-1+b^2}\right\},\\
C_{1,2}^{3,0}=\left\{\frac{b^3 \left(1+b^2\right) \left(b+b^3+4 P\right) \left(-1+b^2+2 b P\right)}{2-3 b^2+b^4}\right\},\\
C_{1,2}^{\{2,1\},0}=\left\{-\frac{4 \left(1+b^2\right)^2 (1+2 b P) (-1+2 b (b+P))}{2-5 b^2+2 b^4}\right\},\\
C_{1,2}^{1^3,0}=\left\{\frac{4 b \left(1+b^2\right) P (1+2 b P)}{1-3 b^2+2 b^4}\right\},\\
C_{1,1^2}^{2,1}=\left\{-\frac{4 b^4 \left(1+b^2\right) P (b+2 P)}{-1+b^2}\right\},\\
C_{1,1^2}^{1^2,1}=\left\{-\frac{2 b \left(1+b^2\right) (b+2 P) \left(-1+b^2-2 b P\right)}{-1+b^2}\right\},\\
C_{1,1^2}^{3,0}=\left\{\frac{4 b^6 \left(1+b^2\right) P (b+2 P)}{2-3 b^2+b^4}\right\},\\
C_{1,1^2}^{\{2,1\},0}=\left\{\frac{4 b^3 \left(1+b^2\right)^2 (b+2 P) \left(-2+b^2-2 b P\right)}{2-5 b^2+2 b^4}\right\},\\
C_{1,1^2}^{1^3,0}=\left\{-\frac{\left(1+b^2\right) \left(-1+b^2-2 b P\right) \left(1+b^2+4 b^3 P\right)}{1-3 b^2+2 b^4}\right\},\\
C_{0,3}^{1,2}=\left\{-6 b \left(1+b^2\right) P (-1+2 b P)\right\},\\
C_{0,3}^{2,1}=\left\{\frac{6 b \left(1+b^2\right) P \left(-4+b^2+b^4+4 b P\right)}{-1+b^2}\right\},\\
C_{0,3}^{1^2,1}=\left\{-\frac{12 b \left(1+b^2\right) P (-1+2 b P)}{-1+b^2}\right\},\\
C_{0,3}^{3,0}=\left\{-\frac{\left(1+b^2\right) \left(12+b \left(b
\left(1+b^2\right) \left(-8+b^2+b^4\right)+12
\left(-3+b^2+b^4\right) P+24 b
P^2\right)\right)}{2-3 b^2+b^4}\right\},\\
C_{0,3}^{1^3,0}=\left\{-\frac{12 b \left(1+b^2\right) P (-1+2 b P)}{1-3 b^2+2 b^4}\right\},\\
C_{0,3}^{\{2,1\},0}=\left\{\frac{12 b \left(1+b^2\right) P \left(-5+3 b^2+2 b^4+6 b P\right)}{2-5 b^2+2 b^4}\right\},\\
C_{0,1^3}^{1,1^2}=\left\{6 b^2 \left(1+b^2\right) (b-2 P) P\right\},\\
C_{0,1^3}^{2,1}=\left\{-\frac{12 b^4 \left(1+b^2\right) (b-2 P) P}{-1+b^2}\right\},\\
C_{0,1^3}^{1^2,1}=\left\{\frac{6 b \left(1+b^2\right) \left(-1+b^2 (-1+4 b (b-P))\right) P}{-1+b^2}\right\},\\
C_{0,1^3}^{3,0}=\left\{\frac{12 b^6 \left(1+b^2\right) (b-2 P) P}{2-3 b^2+b^4}\right\},\\
C_{0,1^3}^{1^3,0}=\left\{-\frac{1}{1-3 b^2+2 b^4}\left(1+b^2\right) \left(1+b^2 (2+b (12 P+b (-7+4 b (b (-2+3 (b-2 P) (b-P))+3 P))))\right)\right\},\\
C_{0,1^3}^{\{2,1\},0}=\left\{-\frac{12 b^3 \left(1+b^2\right) P \left(-2+b^2 \left(-3+5 b^2-6 b P\right)\right)}{2-5 b^2+2 b^4}\right\},\\
C_{0,\{2,1\}}^{1,1^2}=\left\{\frac{2 b^2 \left(-2+b^2\right) \left(1+b^2\right) (b-2 P) P}{-1+b^2}\right\},\\
C_{0,\{2,1\}}^{1,2}=\left\{-\frac{2 b \left(-1+b^2+2 b^4\right) P (-1+2 b P)}{-1+b^2}\right\},\\
C_{0,\{2,1\}}^{2,1}=\left\{\frac{4 b \left(1+b^2\right)^2 P \left(-1+b^2+2 b P\right)}{-1+b^2}\right\},\\
C_{0,\{2,1\}}^{1^2,1}=\left\{\frac{4 b \left(1+b^2\right)^2 P \left(-1+b^2-2 b P\right)}{-1+b^2}\right\},\\
C_{0,\{2,1\}}^{3,0}=\left\{-\frac{2 b^3 \left(1+b^2\right) P \left(-3+2 b \left(b+b^3+3 P\right)\right)}{2-3 b^2+b^4}\right\},\\
C_{0,\{2,1\}}^{\{2,1\},0}=\left\{-\frac{\left(1+b^2\right) \left(4-17 b^4+4 b^8-2 b \left(4+3 b^2+3 b^4+4 b^6\right) P-36 b^4 P^2\right)}{2-5 b^2+2 b^4}\right\},\\
C_{0,\{2,1\}}^{1^3,0}=\left\{\frac{2 b \left(1+b^2\right) \left(-2+b^2 (-2+3
b (b-2 P))\right) P}{1-3 b^2+2 b^4}\right\}\)

In the above expressions, 0 labels $\{\varnothing\}$.

\end{document}